\documentclass[sigconf,nonacm]{acmart}

\pdfoutput=1

\usepackage{float}
\usepackage{pifont}

\setcopyright{none}
\acmPrice{}
\acmDOI{}
\acmISBN{}
\acmConference{}{}{}
\acmYear{}
\copyrightyear{}

\renewcommand\footnotetextcopyrightpermission[1]{}

\begin{document}

\title{TBRD: Tesla Authenticated Broadcast Remote ID}

\author{Jason Veara}
\affiliation{%
  \institution{Northeastern University}
  \city{Boston, MA}
  \country{USA}}
\email{veara.j@northeastern.edu}

\author{Manav Jain}
\affiliation{%
  \institution{Northeastern University}
  \city{Boston, MA}
  \country{USA}}
\email{manav.jain@northeastern.edu}

\author{Kyle Moy}
\affiliation{%
  \institution{Northeastern University}
  \city{Boston, MA}
  \country{USA}}
\email{kyle.moy@northeastern.edu}

\author{Aanjhan Ranganathan}
\affiliation{%
  \institution{Northeastern University}
  \city{Boston, MA}
  \country{USA}}
\email{aanjhan.ranganathan@northeastern.edu}

\begin{abstract}
Mysterious sightings of Unmanned Aircraft Systems (UAS) over U.S. military facilities, suburban neighborhoods, and commercial airports have intensified scrutiny of drone activity. 
To increase accountability, the Federal Aviation Administration (FAA) introduced a Remote ID mandate, which requires unmanned aircraft to broadcast their location, the location of the operator, and their identity in real-time. 
However, current standards leave authentication mechanisms unspecified, enabling spoofing, relay, and replay attacks that can undermine surveillance efforts and potentially disrupt UAS-to-UAS coordination in future deployments.
In this paper, we propose TBRD, a practical system for authenticating Remote ID messages in a manner that aligns with existing standards and UAS capabilities. 
TBRD leverages the TESLA protocol and mobile device TEEs, and introduces a verification mechanism to build a lightweight, mission-scoped authentication system that is both computationally efficient and requires a low communication footprint. 
We evaluate the performance of TBRD using both an FAA-requirements compatible proof-of-concept implementation for performance metrics and a simulated 4-drone swarm mission scenario to demonstrate its security guarantees under adversarial conditions.
Our system provides a 50\% reduction in authentication overhead compared to digital signatures and a 100x reduction in computation time.
Our results demonstrate that TBRD can be integrated into current Remote ID infrastructures to provide a scalable, standards-compliant message authentication for both regulatory and operational use cases.

\end{abstract}

\maketitle

\section{Introduction}
In recent years, the rapid proliferation of Unmanned Aircraft Systems (UASs) has raised pressing concerns around safety, accountability, and regulatory enforcement. Reports of unidentified drones operating near critical infrastructure—such as military bases, commercial airports, and densely populated urban areas—have increased significantly, heightening fears of both accidental collisions and deliberate misuse~\cite{muntean_faa_2025,bonavita_foreign_2025,khalil_mysterious_2024,whitaker_drone_2025}. Law enforcement and airspace regulators face difficulty identifying and holding remote pilots accountable, especially when drone operations lack reliable mechanisms for real-time identification. In response, the U.S. Federal Aviation Administration (FAA) introduced legislation requiring drones to broadcast their location, operator information, and unique identifier in real time, a mandate commonly referred to as the Remote ID rule~\cite{federal_aviation_administration_faa_2019}.

Despite its regulatory importance, the Remote ID standard remains critically incomplete from a security perspective. Most notably, Remote ID broadcasts are unauthenticated, meaning any observer has no way to determine whether a message truly originated from a legitimate UAS. Both Remote ID and Automated Dependent Surveillance-Broadcast (ADS-B) exemplify a fundamental trade-off in aviation safety systems where the priority on open, accessible broadcasts for collision avoidance and airspace awareness inherently creates security vulnerabilities~\cite{khan_survey_2025}. Adding authentication or encryption increases complexity, can reduce interoperability, and potentially compromise the real-time monitoring capabilities that make these systems effective for their primary safety mission. This opens the door to spoofing, relay, and replay attacks that can be executed using simple off-the-shelf tools~\cite{jet_jjshootsremoteidspoofer_2025,tsuboi_drone_2024,muller_cyber-defence-campusdroneremoteidspoofer_2025}. For example, an adversary could replay previously captured broadcasts near a restricted area, triggering false alarms or framing legitimate operators. Such vulnerabilities undermine the integrity of Remote ID data, erode trust in enforcement systems, and, in the future, may even interfere with UAS-to-UAS coordination in decentralized swarm operations.

While digital signatures offer a natural approach to authenticating broadcasts, they introduce practical challenges in UAS deployments. Public key cryptography produces large signature outputs, adding nontrivial overhead to Remote ID messages that are already size-constrained by broadcast standards. Frequent signing also imposes computational burdens on resource-limited UAS hardware. Large-scale key management, revocation, and distribution further complicate deployment, even with assistance from a backend service. In contrast, the Timed-Efficient Stream Loss-Tolerant Authentication (TESLA)~\cite{perrig_tesla_2002} protocol provides similar security guarantees using symmetric cryptography and delayed key disclosure, significantly reducing both computational and communication costs. These properties make TESLA especially suitable for intermittent, real-time systems like Remote ID.

TESLA has been adopted in systems like Galileo GNSS~\cite{european_gnss_supervisory_authority_galileo_2022}. However, applying it to Remote ID requires a fundamentally different design. GNSS satellites are few in number, their identities and codes are fixed, and their public keys rarely change; new satellites are added only every few years. In contrast, drones are purchased, deployed, and retired dynamically, with new Remote ID transmitters appearing unpredictably. Remote ID authentication must therefore accommodate an open and evolving population of devices while ensuring that observers can validate messages from unknown transmitters without prior provisioning. TBRD addresses this challenge by enabling verifiable broadcasts from previously unknown UASs. It utilizes delayed key disclosure and public commitments stored at the USS (UAS Service Supplier), allowing the observers to authenticate messages offline once the corresponding keys become available.

In this paper, we propose TBRD, a system for authenticated Remote ID broadcasting that is compatible with existing standards and realistic deployment constraints. TBRD leverages TESLA for efficient broadcast authentication, uses Trusted Execution Environments (TEEs) on mobile devices to securely manage keychains, and introduces a UAS Service Supplier (USS) component to manage key commitments and enable delayed key validation. The use of TESLA also removes the need for continuous connectivity to a backend: observers can passively collect broadcasts and verify them later once keys are disclosed. TEEs are increasingly present in commercial mobile phones. In cases where they are unavailable, our architecture gracefully falls back to secure local key storage. This flexibility enables TBRD to scale with the evolving drone ecosystem while maintaining low overhead and high verifiability.

Specifically, we make the following contributions. We design TBRD, a Remote ID authentication system that utilizes the TESLA broadcast authentication protocol within the drone ecosystem, enabling verifiable, spoof-resistant broadcasts with minimal computational and communication overhead. We show how mission-scoped keychains can be securely generated and managed using TEEs on commodity mobile devices, reducing the risk of key compromise and enabling deployment on resource-constrained UAS platforms. We integrate a UAS Service Supplier (USS) into the authentication workflow to distribute key commitments and mission metadata, allowing observers to authenticate messages using TESLA-disclosed keys without prior provisioning or persistent connectivity. Our implementation, built on OpenDroneID and Android TEEs, adheres to FAA standards. We compare TBRD against digital signature-based approaches in terms of signing cost, message size, and energy consumption. Finally, using a custom high-fidelity 4-UAS swarm simulation with a decentralized collision avoidance algorithm, we show how TBRD mitigates spoofing, replay, and relay attacks that can otherwise disrupt coordinated flight. Our system provides a 50 percent reduction in authentication overhead compared to digital signatures and requires a fraction of the computational effort — approximately a 100× reduction in signing time from 1.2 seconds to 10 milliseconds. 

\section{Background}
14 CFR Part 107 establishes rules for UAS operations, remote pilot certification requirements, and rules for flying UASs over human beings~\cite{federal_aviation_administration_small_2016}. To ensure remote pilots are operating UASs safely, the FAA now enforces 14 CFR Part 89, which requires operators to comply with the rule on Remote ID.

Pilots can comply with Remote ID in 3 different ways. First, a pilot can operate a UAS that has been equipped with the capability to broadcast identification and location information by the manufacturer. Second, an operator can modify their existing UAS to add the ability to broadcast Remote ID information. Finally, an operator flying within an FAA-recognized identification area (FRIA) satisfies the Remote ID requirement without needing to broadcast Remote ID information. In either of the first two scenarios, the UAS ID, the operator's position, and the UAS's position and velocity information must be broadcast~\cite{federal_aviation_administration_remote_2021}.

The regulations on Remote ID do not dictate the system design of a Remote ID transponder, nor do they require any means of message authentication. It only specifies the functional and performance requirements that any system must meet. In other words, the actual design and development of a Remote ID transponder is left to UAS manufacturers and, in the case of open-source solutions, UAS operators. The UAS manufacturer DJI produces systems that are Remote ID compliant through the use of proprietary protocols called OcuSync. This capability has been included even before the rule on Remote ID was codified into US law. While it remains closed-source, legacy versions of their system, called DroneID, have been reverse-engineered to uncover significant security flaws~\cite{schiller_drone_2023}. Due to the complexities of reverse engineering proprietary implementations, our work aims to enhance open source solutions for Remote ID compliance.

One such effort to promote an open source solution for Remote ID is the publication of the Standard Specification for Remote ID and Tracking, F3411-22a, by the American Society for Testing and Materials (ASTM)~\cite{f38_committee_specification_2022}. This specification provides the technical details that standardize manufacturer approaches and open source efforts to build Remote ID-compliant systems. The specification outlines an optional digital signature-based authentication mechanism. However, the approach requires \textit{persistent} internet connectivity for observers to contact a Broadcast Authentication Verifier Service in order to determine message validity. The main open source effort that complies with ASTM F3411-22a is the project OpenDroneID~\cite{open_drone_id_open_2025}. In addition to offering an open-source C library that individual developers can use to build Remote ID systems, other open-source UAS ground station control developers, such as ArduPilot and MAVLink, offer tools and software to support OpenDroneID integration. This has enabled several manufacturers to build and offer relatively inexpensive add-on modules that can retrofit a UAS with Remote ID capabilities~\cite{holybro_remote_2025,blue_mark_dronebeacon_2025}. Despite the lack of regulations requiring any level of security or the authentication of Remote ID, the ASTM standard outlines a mechanism for broadcasting authentication messsage data. The specifics of the ASTM message structure will be discussed in greater depth in section~\ref{Design}. 

\textbf{The TESLA protocol:} Broadcast authentication has proved to be a complex issue to solve for several years. Many public services, such as the Global Positioning Service (GPS), are broadcast via some wireless medium. These services are vulnerable to growing threats of jamming, spoofing, or record-and-replay attacks. As a result, many works have attempted to provide an efficient means for receivers to verify and authenticate broadcast transmissions. One such method is TESLA~\cite{perrig_tesla_2002}. The motivating concept of the TESLA protocol is that each broadcast packet includes a Message Authentication Code (MAC) computed with a key known only to the transmitter. That key is disclosed at a later time interval, whereby receivers can authenticate previously broadcast messages. TESLA authentication is currently in use by the Galileo global navigation satellite system.  The Galileo Open Service Navigation Message Authentication (OSNMA) officially became operational in July of 2025 and utilizes a combination of the TESLA protocol and digital signature to authenticate satellite positioning messages~\cite{european_spce_agency_galileo_2021, european_gnss_supervisory_authority_galileo_2022}. Furthermore, the TESLA authentication protocol has been proposed as a means to authenticate various types of broadcast services, including Automated Identification System (AIS) for marine traffic information broadcast~\cite{sciancalepore_auth-ais_2022}, as well as  Automated Dependent Surveillance Broadcast (ADS-B) systems for air traffic information broadcast~\cite{berthier_sat_2017}. More recently, it was identified as a potential solution for providing authentication for the very-high-frequency data exchange system (VDES) that is anticipated to enhance a wide variety of maritime communications capabilities ~\cite{wimpenny_pragmatic_2025}. We are the first to propose using TESLA for Remote ID broadcast authentication.

\textbf{TEE capabilities:} TEEs in mobile computing are secure, isolated environments where the most critical operations related to key generation and storage can be safely executed~\cite{jauernig_trusted_2020}. Many mobile applications are becoming increasingly reliant on the services provided by TEEs. They are used to provide services ranging from Mobile Network Operator (MNO) billing, digital rights management, and biometric authentication~\cite{paju_sok_2023,ahmad_enhancing_2013}. In our specific application, the TEE's integrity and secure storage capabilities are leveraged to conduct the transmission setup procedures required by the TESLA protocol.

\section{Design of TBRD} \label{Design}
In this section, we discuss the threat model and assumptions that motivate the design of TBRD.  We also expand upon the conceptual protocol and system architecture.

\subsection{Threat Model and Assumptions} \label{Threat}
Our threat model addresses adversaries who have radio access to Remote ID broadcasts and aim to exploit the absence of authentication to inject false or misleading Remote ID broadcasts. We assume an attacker can either passively observe and/or actively transmit wireless data with basic SDR capabilities (e.g., USRP or HackRF). Furthermore, they have sufficient knowledge to record and replay Wi-Fi beacons or Bluetooth advertisements, the two FAA-approved broadcast media. We also consider attackers who can craft and transmit arbitrary Remote ID messages using open-source toolchains, including those that comply with the Remote ID standard but contain spoofed data. For example, an attacker could capture Remote ID broadcasts from a legitimate UAS mission and replay them at a later time to create the appearance of unauthorized activity. During FAA-designated spectator or sporting events, temporary flight restrictions may be in place over stadiums and other venues that normally fall in open airspace. The attacker replays previously captured Remote ID messages in the restricted zone, making it appear as though a registered operator is conducting an illegal flight. This can trigger enforcement actions or initiate investigations against legitimate operators. In coordinated attacks, repeated replays across multiple zones could be used to overwhelm response efforts and misdirect law enforcement attention.

\begin{figure}
    \centering
    \includegraphics[width=1\linewidth]{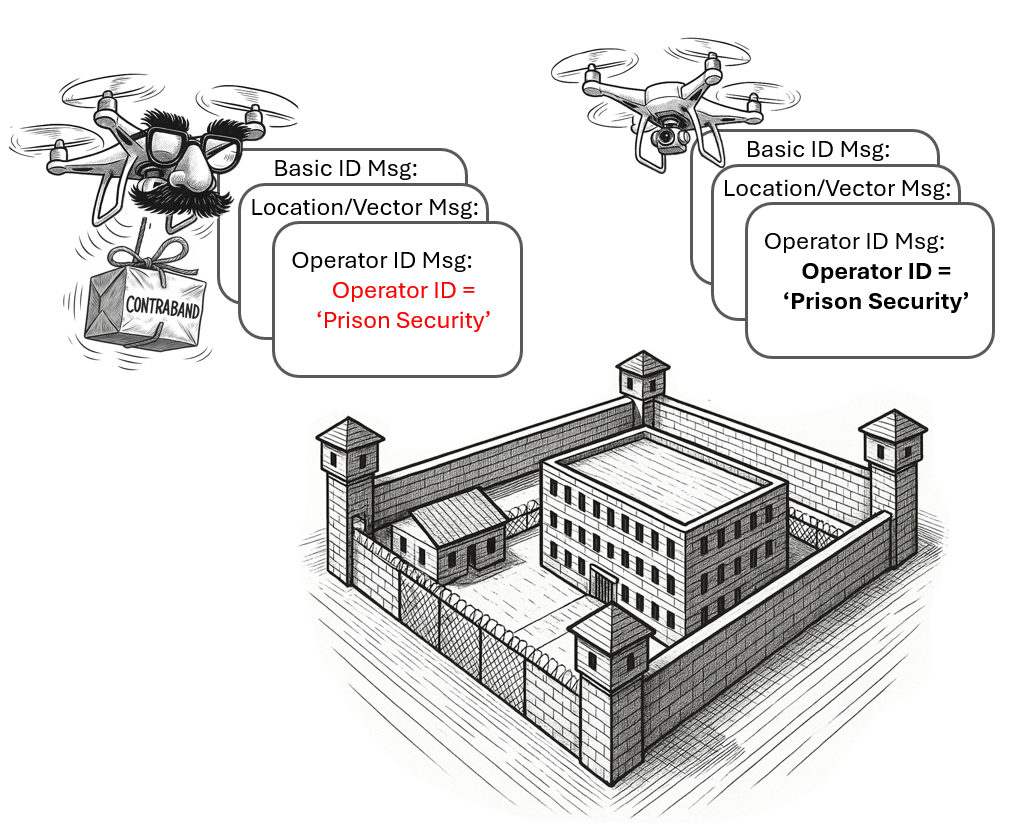}
    \caption{An attacker intercepts Remote ID broadcasts from a legitimate UAS to include Basic ID, Location/Vector, and Operator ID messages. The attacker uses these legitimate transmissions to mask its digital identity in an effort to confuse and disrupt counter-UAS operations.}
    \label{fig:Spoofing}
    \Description[Spoofed Remote ID Attack]{}
\end{figure}

We assume that the adversary does not have access to the internals of the victim UAS (e.g., its firmware, TEE, or secure keys), but may capture a UAS mid-flight and extract mission secrets such as keys or metadata from insecure flash storage. We assume the attacker cannot break standard cryptographic primitives, such as HMAC-SHA256 or AES. The system clocks on both the UAS and the receiver are assumed to be loosely synchronized via GNSS, which is a requirement already specified in the Remote ID specification. We do not consider GPS spoofing or sensor-layer attacks on the UAS itself, as these fall outside the scope of message authentication and delve into sensor integrity and flight controller security. We also assume that the USS is a trusted entity operating under the FAA’s supervision and not colluding with an attacker.

Consider a scenario depicted in figure \ref{fig:Spoofing} where an adversary wishes to operate a contraband-filled drone into a restricted airspace (e.g., prison) monitored by an observer, such as a law enforcement Counter UAS (C-UAS) system. The adversary's drone simply broadcasts standard-compliant Remote ID messages using either a fabricated or cloned identifier. Because the current standard leaves authentication mechanisms unspecified, the observing C-UAS system has no way to determine whether a message truly originated from a legitimate UAS without implementing more complex detection mechanisms. This critical vulnerability enables simple spoofing, relay, and replay attacks, which can undermine surveillance efforts by allowing the adversary's drone to appear compliant, potentially delaying or preventing an intercept. 

In contrast, our proposed TBRD system is designed to thwart such spoofing attacks by default. Furthermore, as we will demonstrate, TBRD's architecture provides specific countermeasures even against more advanced threats, such as capture of a UAS and extraction of mission secret keys from insecure flash storage.

\subsection{System Overview}
\begin{figure}
    \centering
    \includegraphics[width=1\linewidth]{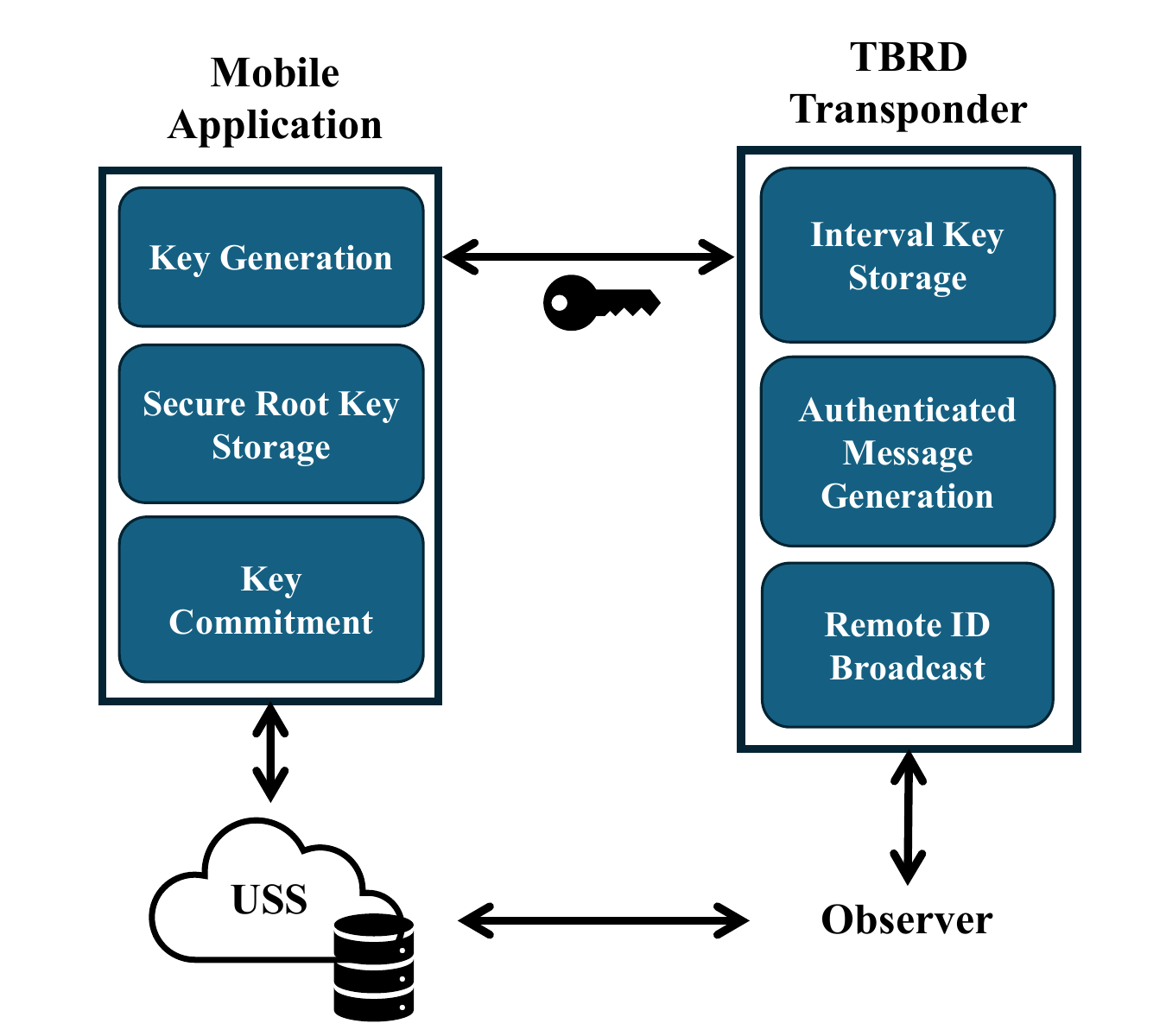}
    \caption{TBRD System Diagram. Components, including the UAS, mobile device, TEE, and USS, and how they interact to secure Remote ID broadcasts.}
    \label{fig:system-diagram}
    \Description[TBRD System Diagram]{TBRD System Diagram. Components, including the UAS, mobile device, TEE, and USS, and how they interact to secure Remote ID broadcasts.}
\end{figure}

Our system is comprised of three major components: the TBRD Transponder (either integrated within the UAS or as an add-on module), a mobile application runing on a device with a TEE, and a trusted USS verification server. Leveraging the TRBD protocol, these components enable UAS pilots to transmit authenticated Remote ID and observers the ability to verify broadcasts. Our system design is depicted in Figure \ref{fig:system-diagram}. 

Communication between the system components will occur according to the following procedure. First, a UAS pilot registers as an operator with the FAA and is assigned an Operator ID, which ties any UAS operations to the individual operator. Furthermore, the operator registers their UAS to receive a unique UAS ID, which identifies the individual UAS. Before the start of the first UAS mission (i.e., a new UAS), the UAS pilot registers the UAS ID (UAS serial number or other number assigned by the manufacturer) with the USS. The UAS pilot’s mobile device is linked to the UAS ID and Operator ID during registration. The commissioning of the new UAS is accomplished over a secure channel. The mobile device becomes, in essence, a physical token that allows the UAS pilot to inform the USS of future UAS missions. Before a UAS mission begins, the pilot configures the UAS's Operator ID, the UAS ID (which can be the UAS's serial number or another number assigned by the manufacturer), and the intended flight duration (i.e., the validity period). The mobile device uses the trusted execution environment to generate all the interval keys, \(K_1,...K_{n-1}\), that will be used to sign TBRD broadcast messages, and the key commitment, \(K_0\), that will be sent to the USS. The key commitment and validity period are transmitted to the USS via a secure communications channel. The USS only accepts mission plans from the mobile devices provisioned during account commission. The interval keys, \(K_1,...K_{n-1}\), are uploaded to the UAS via a secure communication channel. In practice, this could be a secure Bluetooth communication link. The mission start time (defined by the timestamp of the first TBRD message broadcast) is transmitted to the USS. The UAS broadcasts TBRD messages until all interval keys have been exhausted. After all the interval keys are exhausted, the UAS broadcasts in a non-authenticated mode. During the flight, an observer captures at least two TBRD message broadcasts. The interval key disclosure is used to authenticate any previously broadcast message. The observer queries the USS via a secure communications channel to check the validity of the root key, Operator ID, and UAS ID.

\subsection{TESLA-Based Remote ID Authentication}

\begin{figure}
    \centering
    \includegraphics[width=\linewidth]{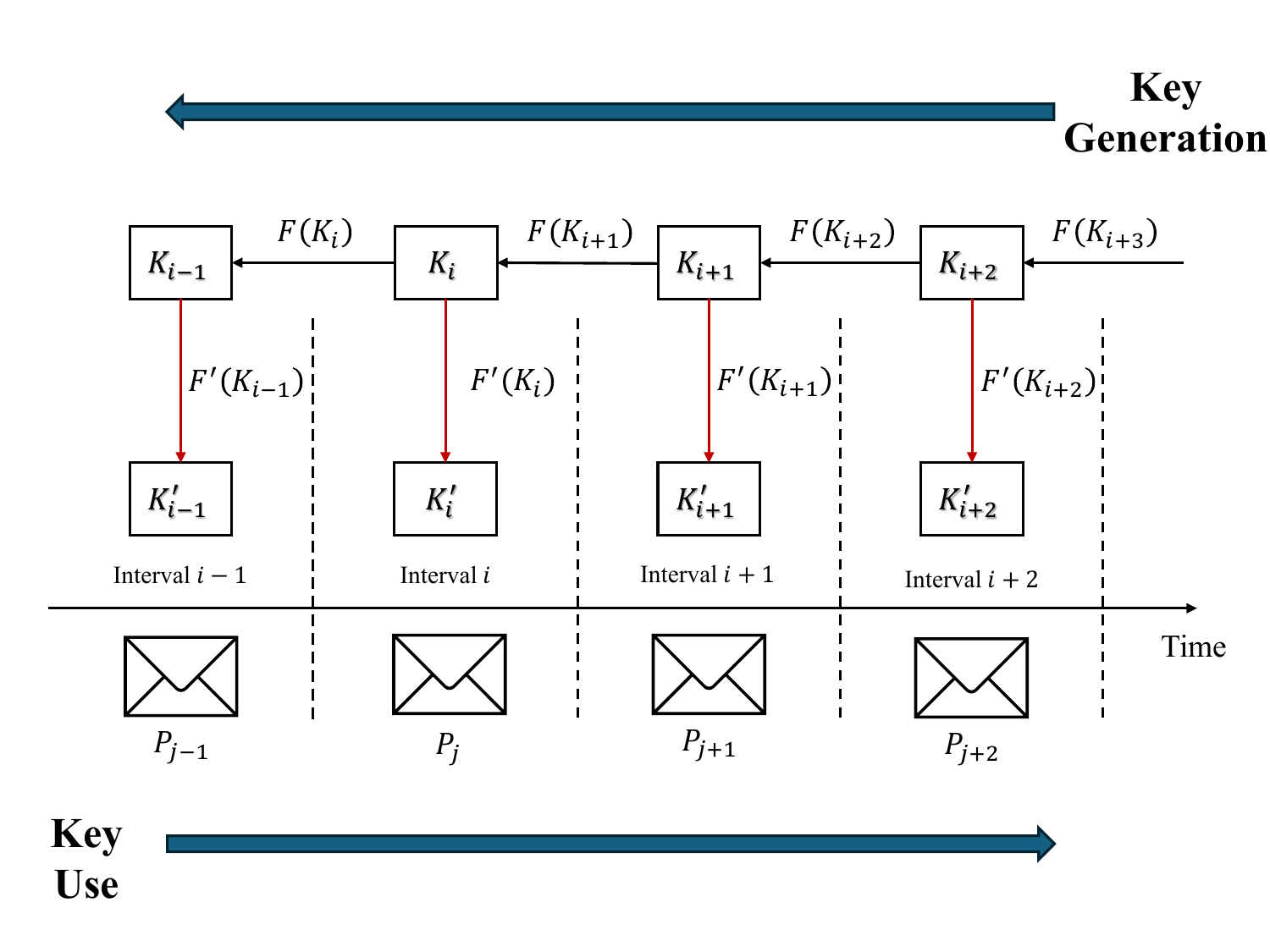}
    \caption{TESLA key chain generation. Keys are generated in reverse order of key disclosure and message authentication.}
    \label{fig:TESLA_Keys}
    \Description[TESLA Background]{TESLA key chain generation. Keys are generated in reverse order of key disclosure and message authentication.}
\end{figure}

The TESLA protocol achieves asymmetric authentication properties, using symmetric authentication keys. Figure \ref{fig:TESLA_Keys} depicts the process of key generation. The sender divides anticipated transmission time into \(n+1\) intervals. The intervals are defined by the index set \(I=\{0,1,2,...n\}\). The sender and receiver are loosely time synchronized, and the receiver is aware of the time intervals in advance. The sender then generates a set of keys using a one-way keychain function. We define the set of keys \(K = \{K_i|i\in I\}\). As a result of the one-way key generation function, knowledge of \(K_{i-1}\) does not allow the computation or recovery of \(K_i\). Additionally, the keys are generated in reverse order of key use, where \(K_n\) is generated first but used last. To prevent the reuse of keys (i.e., using \(K_I\) multiple times), a pseudo-random function is used to generate \(K'_i\), which will be used to generate a message authentication code (MAC). The sender can now begin to generate broadcast messages. The sender generates a message \(M_j\), where \(j\) is the current interval index, \(i\) is the corresponding key interval, and \(d\) is an adjustable disclosure delay. The constructed packet \(P_j\) is defined by equation \ref{eq:TESLA_packet}.

\begin{equation} \label{eq:TESLA_packet}
  P_j=\{M_j||MAC(K'_i,M_j||K_{i-d})\}   
\end{equation}

The disclosure of the previous interval key \(K_{i-d}\) along with the message \(M_j\), allows a receiver to authenticate all previous messages. However, the receiver must wait until the next interval to authenticate the message broadcast in the current interval. A receiver can also generate \(K_0\), to validate that any interval key, \(K_i\), is a valid member of the set \(K\). In this manner, the value \(K_0\) becomes a token to validate the commitment to the set of keys from the one-way keychain. If either the validation of the message itself fails or the interval key is not a member of the one-way keychain, the message cannot be authenticated.

The benefits of the TESLA algorithm that make it particularly well-suited to Remote ID broadcast authentication are as follows: 

\begin{enumerate}
    \item Low computational overhead
    \item Robustness to packet loss,
    \item Scalability.
\end{enumerate}

The computational requirements of the Remote ID transmitter are extremely low. A single SHA-256 hashed MAC, computed once every second, enables us to comply with the FAA Remote ID requirements. Due to the dynamic nature of UAS flight, robustness to packet loss is critically important. The system is designed so that any observer within the UAS's line of sight range can receive and authenticate broadcast Remote ID data. The loss of any single packet will delay the observer's ability to verify the previously received message. Once any subsequent message is received, all previously recorded messages can be verified.

For TESLA to operate correctly, several conditions must be met. First, as mentioned previously, the sending and receiving entities must have loose time synchronization. In other words, the receiver must know the upper bound of the difference between the sender's and receiver's clocks. The key interval must be chosen such that it takes into account this error. Second, the receiver must know the key disclosure schedule, including the interval duration, \(T_{int}\), key disclosure delay \(d\), and a key commitment to the keychain, \(K_i\). Section \ref{Implementation} demonstrates how TBRD fulfills these requirements by utilizing tightly synchronized GPS time, Wi-Fi beacon broadcasts, standardized key generation via the TEE of a mobile device, and verification of the key commitment through a trusted third party.

TESLA prevents manipulation of message data in transit, as the key used at transmission time is only known to the broadcasting entity (in our case, the UAS). It does so in a manner that is robust to packet loss with relatively low computational overhead. Furthermore, the disclosure of keys is broadcast to all parties. No previous key exchange must occur for an observer to verify a message.

\subsection{Key Management via Mobile TEE}
As outlined above, the interval keys used by the TESLA protocol are generated from a one-way keychain where the input of a single value is hashed iteratively. Generating and storing this key chain's "seed" material is critical to the system's security. If the seed material for the keychain is exposed to an attacker at any point, the entire chain of keys can no longer be trusted. We propose using a mobile device TEE to generate and store the seed key securely.

In the setup phase, the sender divides the anticipated broadcast time into uniform duration intervals \(T_{int}\). Based on the interval duration and total length of the mission, the sender can generate \(n+1\) keys where:

\begin{equation} \label{eq:Mission Time}
    n=\frac{Total Flight Time}{T_{int}}
\end{equation}

Once the length of \(n\) is determined, the TEE securely generates a random \(K_n\) and derives the interval keys. The set of interval keys \(K_1,...K_{n-1}\) can then be uploaded securely to the UAS. Key \(K_n\) is stored securely within the TEE and is never disclosed to other entities. However, the last key generated in the one-way key derivation process \(K_0\) is used to commit to using the derived interval keys for a specific mission. This process is repeated whenever an operator decides to fly a mission. A new \(K_n\) is generated for each mission, and the same values are never reused. The long-term security of the system is ensured as long as the interval keys can be securely shared with the UAS, the TEE produces the interval keys securely, and \(K_0\) is never disclosed. Should a UAS ever be captured, the only compromised key material is the interval keys that have not yet been disclosed during normal operations. 

\subsection{Observer Verification via USS}
To satisfy the remaining security requirements, we need to provide a mechanism for an observer to verify the keychain in use. Additionally, we must protect our system from an attacker who simply replays broadcasts from an old mission. In addition to the attacks outlined in \ref{Threat}, there are many other scenarios where an attacker may want an observer to receive conflicting or incorrect Remote ID information. Therefore, we must provide additional system functionality that connects the intended mission with the observer to verify flight data. We propose the following.

Once the mobile device has generated the set of keys, the key commitment value, and mission start, they are published to a trusted server. Once the mission is complete, the mobile device sends the mission end time. The ASTM standard proposes the use of a UAV Service Supplier (USS) to facilitate the coordination and distribution of network Remote ID ~\cite{f38_committee_specification_2022}. In network Remote ID, the operator publishes Remote ID information to a USS via a \textit{persistent} internet connection from a ground control station (i.e., cellular or Wi-Fi). We adopt this concept of the trusted USS into our system design, however the internet connection no longer must be \textit{persistent}. When an observer receives a TBRD-authenticated broadcast from a UAS, the observer first verifies the authenticity of the received packet using the follow-on key disclosure. The USS can then provide the mission's pre-disclosed key commitment value, start, and end time. We assume that all these transactions occur over cryptographically secure connections. If the observer receives a message but cannot immediately query the USS, this verification process can be accomplished at a later date. Additional assumptions and specific verification algorithms will be further discussed in section \ref{Implementation}.

\section{Implementation} \label{Implementation}
We implemented our design by leveraging the existing protocols defined by ASTM F3411-22a and the open-source code base OpenDroneID ~\cite{open_drone_id_open_2025}. OpenDroneID is a C-based library that encodes and decodes OpenDroneID messages, which can be transmitted/received over a wireless interface, such as Wi-Fi. To facilitate rapid prototyping and easier integration with existing simulation infrastructure, we implemented our system using Python. Our source code for a proof-of-concept implementation can be found \href{https://github.com/t-brd/tbrd}{here}. The implementation includes a mobile application, a Linux-based transmitter, receiver, and USS server.

\begin{figure}
    \centering
    \includegraphics[width=1\linewidth]{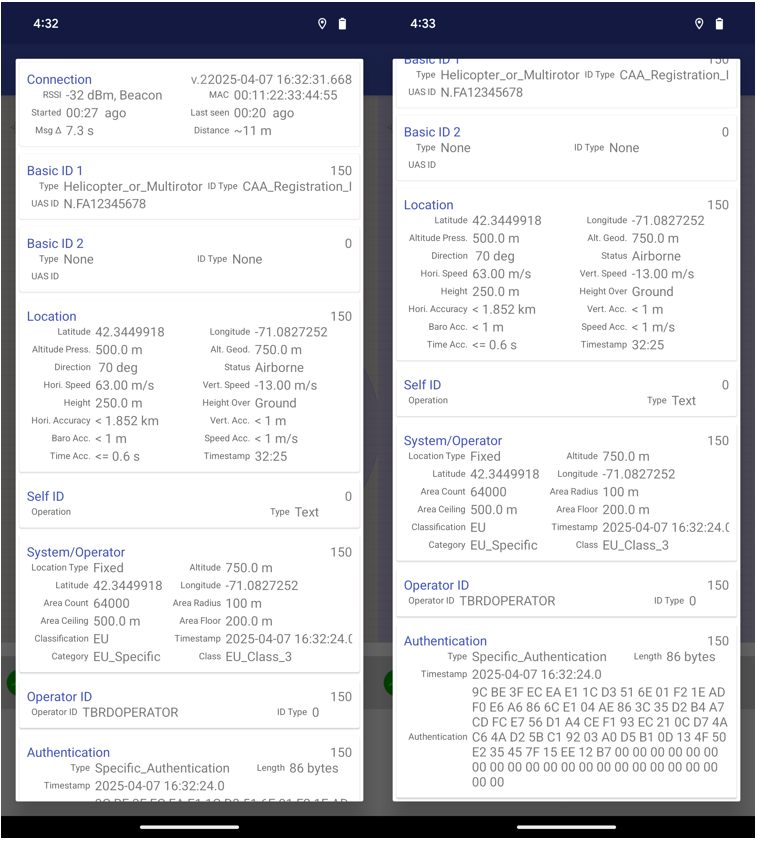}
    \caption{OpenDroneID application displaying the authentication messages transmitted by our proposed system, demonstrating successful message reception and validation on standard Remote ID monitoring devices.}
    \label{fig:OpenDroneID_App}
    \Description[DroneScanApp verification]{OpenDroneID application displaying the authentication messages transmitted by our proposed system, demonstrating successful message reception and validation on standard Remote ID monitoring devices.}
\end{figure}

The ASTM standard offers four broadcast transport mechanisms for broadcast Remote ID, primarily driven by the available payload capacity. Those options are Bluetooth Legacy (4.x), Bluetooth 5.x Long Range, Wi-Fi Neighbor Awareness Network (NAN), and Wi-Fi Beacon (vendor-specific information element in the SSID beacon frame). We choose to implement broadcast via Wi-Fi Beacon. Special attention must be paid to selecting a wireless broadcast transport mechanism when utilizing TESLA authentication. Wi-Fi beacons allowed for a key interval large enough to comply with ASTM F3411-22a and maintain security guarantees of the TESLA protocol. We have validated that our system does comply with ASTM F3411-22a utilizing the open source OpenDroneID Android application ~\cite{open_drone_id_opendroneidreceiver-android_2025}. Figure \ref{fig:OpenDroneID_App} shows our TBRD WiFi broadcast successfully received and parsed by the application, demonstrating compatibility with the standard. 

\subsection{Mobile Application and TEE}
The mobile application serves as the primary interface for pilots to initiate and manage UAS missions, while also facilitating secure key generation by utilizing the TEE on the mobile device.~\cite{android_android_2025}

Before commencing a mission, the pilot uses the app to input operational parameters such as the planned mission start time, end time (upon mission completion), UAS ID, and Operator ID. The start and end times are used to calculate the mission duration, which in turn determines the number of TESLA keys needed. All of this information is required in our implementation of TBRD.

The mobile app leverages the TEE to generate a sequence of 32-byte-long keys following the TESLA protocol framework. In the TEE, the key generation process begins with the TEE selecting a random 32-byte string as the seed key, which serves as the input for the SHA-256 hash function to generate subsequent TESLA keys. The TEE produces a chain of \(n + 1\) keys, where \(n\) is defined by Equation \ref{eq:Mission Time}.

When the entire keychain is generated, it is exported from the TEE and uploaded to the UAS via a secure upload mechanism. Next, the app creates a file containing the UAS ID, Operator, ID, mission start time, mission end time, and the \(n+1 \) key. This is sent to the USS securely for TBRD message verification by any observer. In our proof-of-concept implementation, this step is simulated by manually transferring the key file to the TBRD transmitter application.

\subsection{TBRD Broadcast Transmitter}
TBRD is designed to comply with the ASTM F3411-22a standard, ensuring compatibility with any existing hardware that utilizes it. 

Before flight, the interval keys are uploaded to the UAS via a secure communications channel. Implementing a secure upload of keys from the mobile application to the UAS is beyond the scope of this work. There are numerous ways to execute this step securely and efficiently. The interval keys \((K_i)\) and the root Key \((K_0)\) are saved in the transmitter memory. The number of keys and key length must be chosen considering the storage space available on the transmitter.

The TBRD transmitter will populate the Basic ID (Type 0x0), Location/Vector Message (Type 0x1), System Message (Type 0x4), and Operator ID Message (Type 0x5) according to ASTM F34411-22a. All static and dynamic fields are populated via an internal MAVLink communication channel with the flight controller. For testing and evaluation purposes, the TBRD transmitter enables the transmission of a static dataset and can also ingest dynamic flight information from an ArduPilot software-in-the-loop simulation.

The Authentication Payload is generated by concatenating the messages in the order indicated below, along with a 4-byte interval counter, \(i\). The use of an interval counter serves two purposes. First, it prevents message replay by an adversary. Second, the counter, along with the disclosed mission start time, allows the receiver to determine the current broadcast interval and ensure the key remains secret at the time of use. The resulting message is 104 bytes long with the form:
        
        \begin{align}
            \text{Authen}&\text{tication Payload}= \notag\\ &(i \| \text{Basic ID Message} \| \notag\text{Location/Vector Message} \| \notag\\ &\text{System Message} \| \text{Operator ID Message}) \notag\\
        \end{align}
            
The Authentication Message is generated by calculating a SHA-256 hashed message authentication code of the authentication payload using the current interval key \(K_i\). The entire Authentication Message is created by concatenating the following fields, where \(K_{i-d}\) is the previous interval key disclosure. The whole authentication message is 68 bytes in length.
        \begin{align}
            \text{Auth}&\text{entication Message}=\notag\\
            &(i\|\text{HMAC(Authentication Payload},K_i)\|K_{i-d})
        \end{align} 

The first 17 bytes of the Authentication Message are encoded in a “page 0”, Type 0x2 Authentication Message. The remaining 51 bytes are encoded in Type 0x02 Authentication Messages for a total of 4 pages of authentication data.

Finally, a type 0xF Message pack is created using the Basic ID, Location/Vector ID, System Message, Operator ID, and the 4 Authentication messages. The Message pack is then transmitted over the wireless interface once per second to comply with the broadcast rate specified in ASTM F3411-22a.

Because we opted to send our TBRD messages within the vendor information element of a broadcast Wi-Fi beacon, special care must be taken to ensure this implementation does not violate the security requirements of the TESLA protocol. Specifically, we need to ensure that once the broadcast beacon message has been crafted, it can be sent within the intended interval. However, a wireless network defined by IEEE 802.11 may prevent the beacon's broadcast within the interval. Wi-Fi networks implement a carrier sense multiple access with collision avoidance (CSMA/CA) feature, which requires the transmitting station to verify that the channel is idle before transmission \cite{ieee_ieee_2025}.  If the channel is busy, the transmitting station must wait a specified amount of time before attempting transmission. Therefore, the delay may cause the transmission to occur outside the intended interval. The transmission of a message outside its intended interval has profound security implications. For example, a system implements a key disclosure delay of \(d=1\). Message \(M_4\) is a message meant for transmission in interval \(i=4\). If \(M_4\) is delayed and sent in interval \(i=5\) along with Message \(M_5\), the disclosure of the key used to authenticate \(M_4\) occurs in the same interval of key disclosure. This would violate the security guarantees offered by the delayed key disclosure mechanisms. As a result, we must be ready to discard and prevent the transmission of any packets outside the intended interval period.

Controlling the transmission of Wi-Fi beacons within a software implementation of the TBRD protocol requires implementing additional safeguards to manage a potentially dynamic wireless channel effectively. One approach would be to build custom Wi-Fi firmware to prevent beacons from being broadcast outside their intended interval. However, we chose an approach that defined a permitted broadcast period within the key interval. This approach also considers the TESLA protocol enhancements from \cite{jahanian_analysis_2015}, which address the maximum clock skew expected from GPS synchronization errors. 

\begin{figure}
    \centering
    \includegraphics[width=1\linewidth]{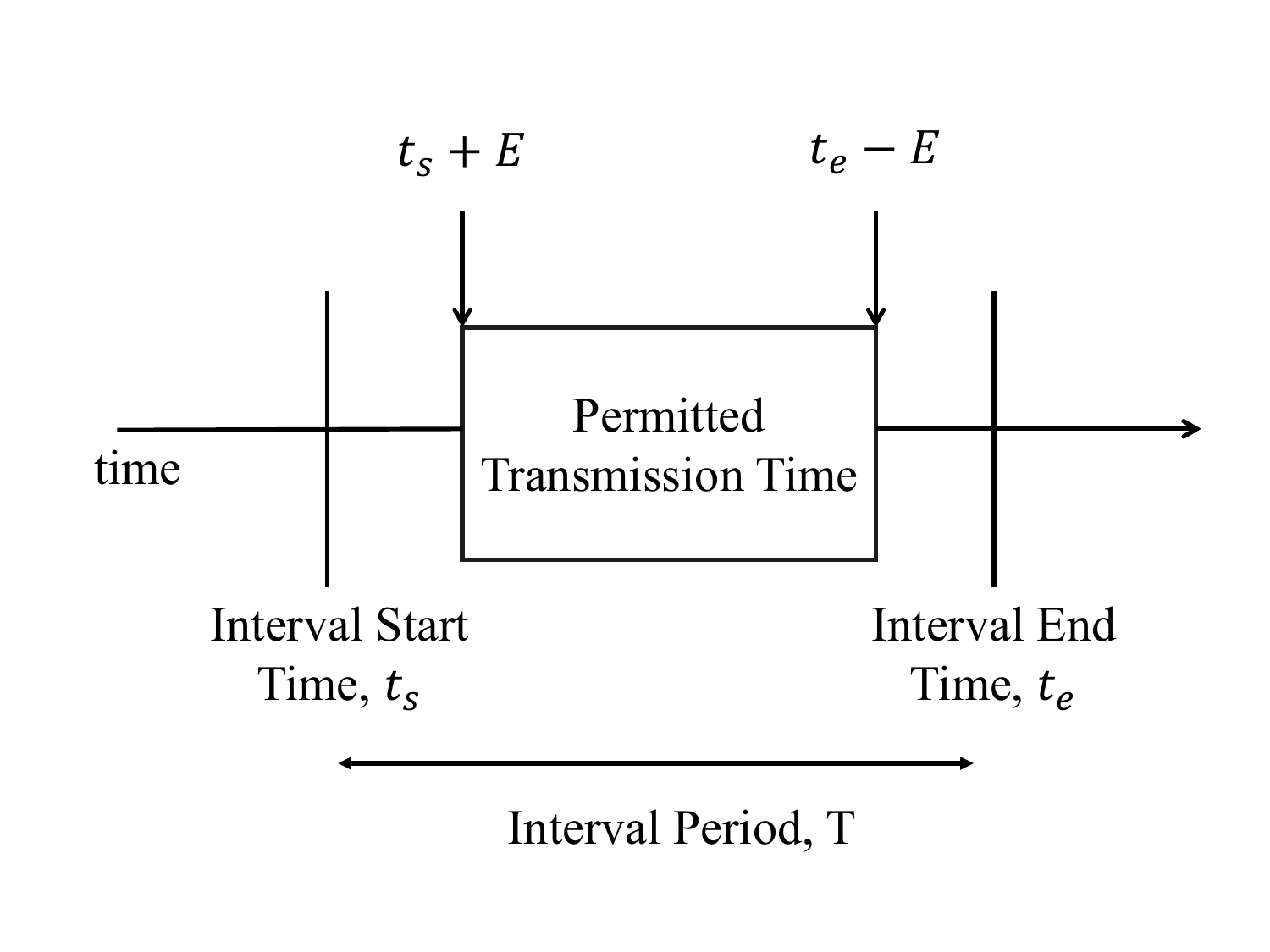}
    \caption{Definition of a permissible transmission window within each key interval. This is the period where Wi-Fi beacon broadcasts are authorized to ensure a balance between system functionality and security requirements.}
    \label{fig:permitted_tx_time}
    \Description[Permitted Tx Time]{Definition of a permissible transmission window within each key interval. This is the period where Wi-Fi beacon broadcasts are authorized to ensure a balance between system functionality and security requirements.}
\end{figure}

As depicted in Figure \ref{fig:permitted_tx_time}, we explicitly define a permissible period within the key interval during which the transmitter can broadcast the beacon. The delay \(E\) identified can be customized and fine-tuned to weigh system functionality and security. 

\subsection{TBRD Receiver}
Once an observer receives a message, it must conduct a series of checks to ensure it is valid. There are two main steps to message validation.  

First, the message is buffered in memory until the time of key disclosure. Since our implementation uses a key disclosure delay equal to 1, the receiver will buffer the message for a single key interval. Once the key is received, the receiver calculates the HMAC of the received authentication payload and checks that it matches the HMAC reported in the authentication message. If the calculated HMAC does not match the received HMAC, then the message must be marked as invalid.

Next, the receiver will send a request to the USS to ensure that: 
\ding{182} the keychain in use by the operator was disclosed correctly and is valid for the observed period of use, and 
\ding{183} the interval keys are being used in the correct interval (i.e, keys are not being reused by an attacker)

The query to the USS includes the observer operator ID, UAS, and time of observation (UTC). If a valid mission exists for that operator within a valid mission period, the USS will respond with the key commitment value \(K_0\) and the mission start time in UTC. With this information, the receiver will proceed to validate the key.

\begin{figure}
    \centering
    \includegraphics[width=\linewidth]{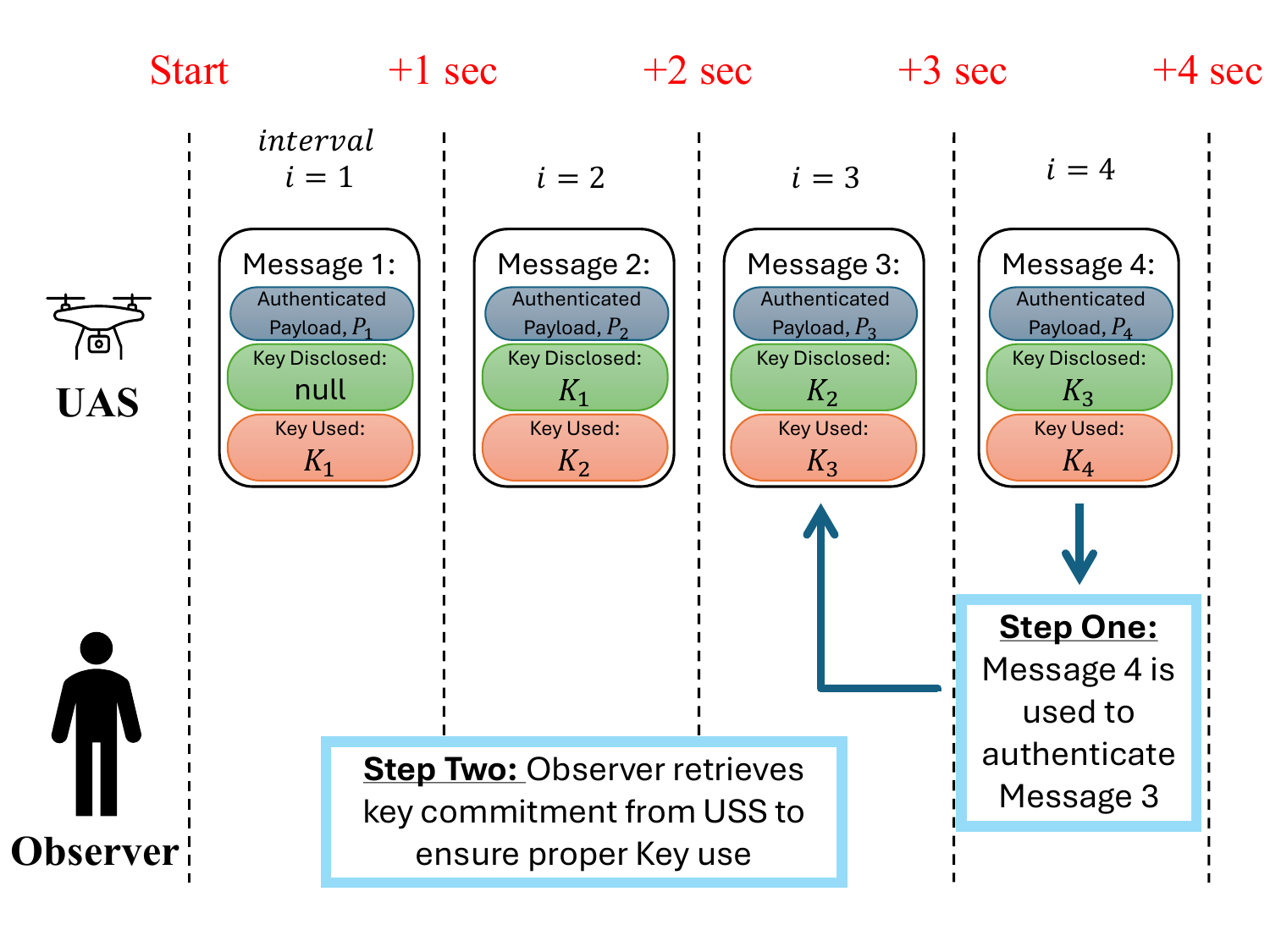}
    \caption{Key verification workflow showing how the observer authenticates received messages through subsequent key disclosure, USS coordination for key-commitment validation, and temporal checks to confirm that keys remained secret during their usage intervals.}
    \label{fig:verification}
    \Description[Verification process]{Key verification workflow showing how the observer authenticates received messages through subsequent key disclosure, USS coordination for key-commitment validation, and temporal checks to confirm that keys remained secret during their usage intervals.}
\end{figure}

Figure \ref{fig:verification} demonstrates a specific scenario where the observer receives a message transmitted at intervals 1 through 4. The interval period \(T\) equals one second, and the disclosure delay is 1.

\begin{enumerate}
    \item The observer receives message \(M_3\) at some time \(t\) and waits for the receipt of \(M_4\). Once received, it uses the key disclosed in \(M_4\) to validate the authenticity of the HMAC broadcast in \(M_3\).
    \item USS Verification:
    \begin{enumerate}
        \item The observer queries the USS, reporting the observer ID, UAS ID, and the time observer \(t\). The USS provides the reported mission start time and key commitment \(K_0\) reported by the observer.
        \item The observer hashes key \(K_3\) which was disclosed in message \(M_4\) a total of \(i = 4\) times to generate the observer key commitment value.  This observed key commitment value is compared against \(K_0\) reported by the USS. If the values match, the observer knows a valid keychain is in use.
        \item The observer subtract \(i=4\) second from \(t\), the arrival time of message \(M_4\). The following equation generalizes this.
        \begin{equation}
        Time_{K_0}=t-i
        \end{equation}
        If: \(Time_{K_0}<Mission StartTime\)
        Then: \(K3\) was still a secret at the time of use, and the observer determines that the key was used in the proper interval
    \end{enumerate}
\end{enumerate}

\subsection{USS Server}
The verification server, or USS, stores mission-specific information to validate TBRD messages. The exact implementation of the server could vary significantly in a production environment. However, the basic functional requirements are that it must be able to communicate with authorized users, store the mission data, and respond to mission data queries from receivers. Our implementation utilizes a simple Python dictionary to store mission data and communicates with the transmitter and receiver using ZeroMQ and Google Protocol Buffers ~\cite{google_protocol_2025, zeromq_zeromq_2025}.

The security of data in transit, user enrollment, and authentication are critical to the overall system. Several solutions exist that could provide the necessary security to protect communications between the mobile device and the production USS server. Our contribution focuses on implementing the TBRD protocol. Therefore, our proof-of-concept implementation assumes that these security controls have already been met.

\section{Security and Performance Analysis} \label{Analysis}
\subsection{Swarm Simulation Testbed}

To evaluate the robustness of the TBRD protocol and our implementation, we simulated an autonomous UAS swarm. In the swarm simulation, each UAS relies on the location information broadcast by Remote ID to make collision avoidance decisions while executing a pre-defined mission. We demonstrate several attacks on broadcast transmission and evaluate the extent to which TBRD can protect from these attacks.

% \subsection{Platform for Robotic Integration and Simulation of Multi-Agent Systems}

We leveraged an internal UAS swarm simulation testbed (name anonymized for review) to enable rapid simulation and prototyping of autonomous UAS swarm algorithms. At the heart of the system is the Gazebo robotics simulator. The Gazebo platform can simulate physical environments, real-world sensors, and communication transport mechanisms. The testbed utilizes Docker to simulate a swarm network, connecting a set of containerized ArduPilot flight controllers \cite{docker_docker_2025}. ArduPilot is an open-source autopilot solution that can simulate various types of autonomous vehicles ~\cite{ardupilot_ardupilot_2025}. A current limitation of the testbed platform is the inability to simulate a wireless communications channel. As a result, we chose to implement communications over the Docker network where the contents of the WiFi beacons were broadcast via UDP. While this does not perfectly emulate a wireless channel that must comply with the 802.11 protocol, the connectionless nature of UDP broadcast allows us to evaluate the critical features of the TBRD system. With these combined features, the testbed can emulate a fully decentralized UAS swarm architecture capable of executing complex mission scenarios. 

\begin{figure*}[!t]
    \centering
    \includegraphics[width=\textwidth]{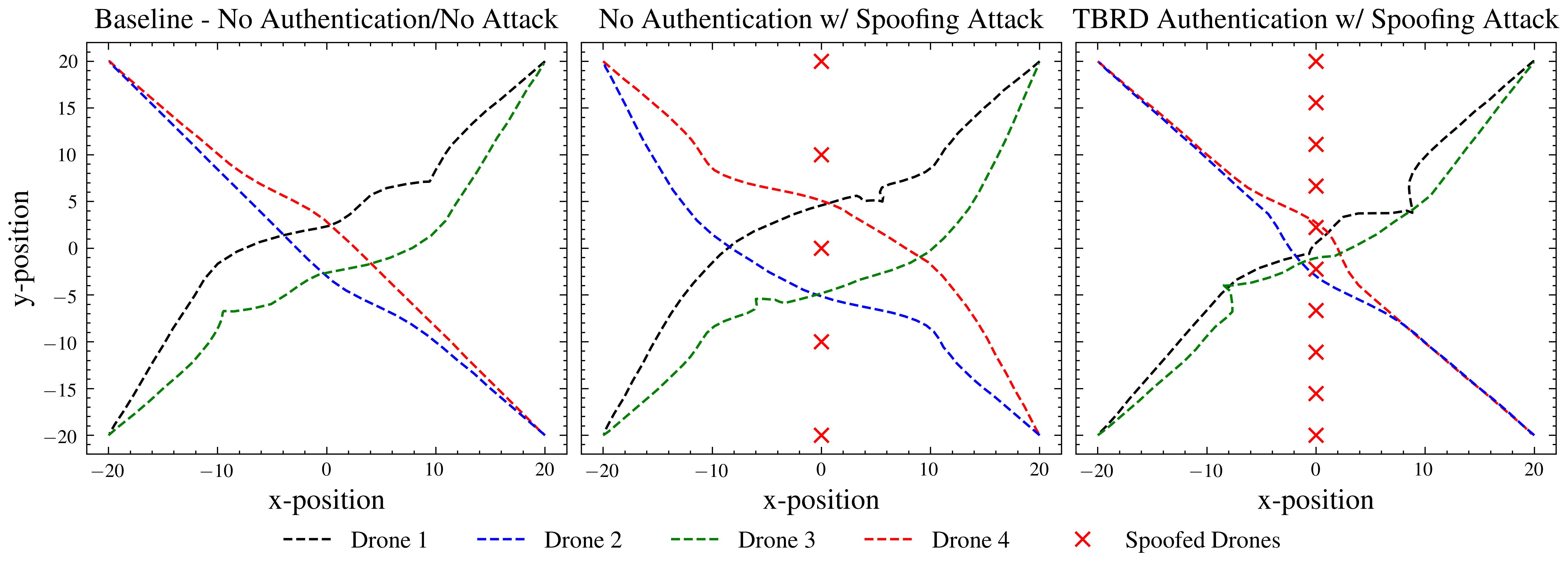}
    \caption{This figure demonstrates the positive impact of TBRD Authentication on collision avoidance for swarms. The first graph indicates the path traveled by four drones during a baseline scenario of using unauthenticated Remote ID.  The second graph illustrates how the same four drones will adjust their paths to avoid spoofed drones.  The last graph shows that the four drones can successfully detect spoofed drones when using TBRD Authentication and complete their mission without interruption. }
    \label{fig:Evaluation_Scenarios}
    \Description[TBRD Evaluation]{TBRD Evaluation Scenarios Results. This figure illustrates how the UAS swarm moves without Remote ID, with no attack, and with Remote ID authentication spoofing a line of UASs, as well as TBRD authentication with Remote ID spoofing UASs.}
\end{figure*}

\textbf{Collision Avoidance Algorithm:} One challenge that autonomous UAS swarms must overcome is executing a given mission or task without collision between individual agents of the swarm. TBRD provides a robust means of broadcasting position, with an added layer of authentication to ensure only trusted locations are used for collision avoidance calculations. 

Once the locations of surrounding agents in the swarm have been collected, some form of algorithm or procedure must process that information to decide how to act. Evaluating the relative performance of differing collision avoidance algorithms is beyond the scope of this investigation. Instead, we aimed to evaluate the relative safeguards against attacks against communications-based ranging when using TBRD. As such, we evaluated one such potential collision avoidance algorithm that provides decentralized decision-making capabilities to our swarm simulation: optimal reciprocal collision avoidance (ORCA) ~\cite{siciliano_reciprocal_2011}. The availability of code implementation weighed heavily on our decision to use ORCA as our collision avoidance decision algorithm ~\cite{nazecic-andrlon_muonpyorca_2025}. ORCA uses the concept of velocity obstacles to derive collision-free conditions for a set of system parameters~\cite{siciliano_reciprocal_2011}. 

\textbf{Experimental Setup:} Our experiment consisted of a simulation of 4 UASs, each beginning their mission on the corner of a 40-meter by 40-meter square. Each UAS is instructed to take off, fly to an altitude of 5 meters, and then fly to the opposing corner of the square, landing upon arrival. This high-level mission layer directive is augmented with the ORCA collision avoidance mechanism, whereby each UAS should take into account the location of every other UAS in the swarm and avoid collision throughout the mission. Our implementations of the ORCA algorithm used a timestep of 1 sec, a radius value of 2 meters, and a \(\tau\) of 5 seconds. We first developed a baseline mission using Remote ID to share location information.  We then simulated two additional scenarios, one using Remote ID with no authentication, and another using TBRD to share location.  During the final two scenarios, we also simulated a spoofing attack where a line of UASs was placed along the y-axis. This simulates a scenario where an attacker has knowledge of the intended mission scenario and is within close proximity to inject fake Remote ID broadcasts.

\textbf{Results:}
The results of our experimentation are summarized in Figure \ref{fig:Evaluation_Scenarios}. Video recordings of each simulation scenario are also located on our \href{https://sites.google.com/view/tbrd/tbrd}{website}.  First, the baseline scenario demonstrates that we can correctly implement broadcast Remote ID communications over our simulated wireless channel. Remote ID proved to be a sufficient mechanism to share location information among members of the UAS swarm. Furthermore, our implementation of the ORCA collision avoidance algorithm prevented collisions among individual swarm agents.

The attack launched during the second scenario consisted of five spoofed UASs, strategically placed in the path of the UAS swarm agents. Since no mechanism exists to identify the true nature of the spoofed UASs, the swarm integrated these Remote ID broadcasts into the collision avoidance solution. There were no collisions among the swarm agents, and they collectively avoided collision with the spoofed UASs. However, the presence of the spoofed UASs did alter the path of every agent in the swarm. While the mission was still ultimately a success, this demonstrates the feasibility of a spoofing attack that can change the performance and behavior of a decentralized swarm.

Finally, the last scenario demonstrates how TBRD can effectively protect the UAS swarm from ingesting false information in collision avoidance calculations.  The number of spoofed UASs in this attack is increased from 5 to 10 to stress the performance of the ORCA collision algorithm and the authentication procedures executed by TBRD. The attacker in this scenario has complete knowledge of the TBRD protocol and can generate a set of valid interval keys for use during TESLA authentication.  The attacker intends to maintain a stealthy profile. Despite their knowledge of the TBRD protocol, they are unable to properly register their mission data and key commitment with the USS. As a result, the verification checks flag the UASs as invalid, and the UASs do not alter velocity as they approach the spoofed location.

\subsection{Spoofing-Replay-Timing Attack Resilience}

\begin{table*}[!t]
\centering
\caption{Comparative Security and Performance Analysis of Remote ID Authentication Approaches}
\label{tab:security-comparison}
\begin{tabular}{l|c|c|c}
\hline
\textbf{Security Property} & \textbf{No Authentication} & \textbf{Digital Signatures} & \textbf{TBRD} \\
\hline
\hline
Spoofing Resistance & \ding{55} & \checkmark & \checkmark \\
\hline
Computation time & N/A & High (1.2 sec) & \textbf{Low (10 ms)} \\
\hline
UAS compromise & N/A & \ding{55} long term private key compromise & \checkmark key compromise only for current mission\\
\hline
Packet Loss Tolerance & N/A & Poor & \textbf{Excellent} \\
\hline
Real-time Verification & N/A & Difficult & \textbf{Easy} \\
\hline
Key Management Complexity & None & High & \textbf{Low} \\
\hline
Authentication Overhead & 0 bytes & \(>\)139 bytes & \textbf{68 bytes} \\
\hline
Scalability (observers) & N/A & Limited & \textbf{Unlimited} \\
\hline
\end{tabular}
\end{table*}

As the swarm simulation demonstrated, TBRD's USS verification mechanism is fully resilient against mission-level spoofing attacks. An adversary broadcasting messages without a valid, USS-registered key commitment will be immediately identified.
In this section, we analyze more subtle attacks that target the intrinsic properties of the TESLA protocol itself and its impact on TBRD security guarantees: (1) the temporary window of uncertainty created by delayed key disclosure, and (2) attacks on the system's time synchronization. The TESLA protocol is a key feature of TBRD that protects against attacks during active missions. The experimental results presented in this section demonstrate the security guarantees provided by the TBRD protocol.  However, two potential weaknesses that can be targeted to affect the security guarantees of TESLA are the key disclosure period and time synchronization. 

The primary trade-off of TESLA is the key disclosure period, which creates a temporary window of uncertainty. An attacker can inject a fake message during the current key interval.  The observer receives this message but cannot validate or reject it until the next interval when the corresponding key is disclosed.
While the attack is detected (and the fake message is ultimately discarded) upon key disclosure, the system is blind to the forgery for the duration of that interval. Care must be taken not to trust received data implicitly during this window. For highly sensitive, real-time applications such as collision avoidance, a 1-second delay in verification is a critical operational consideration that system designers must account for and configure according to their specific use case.

\begin{figure}
    \centering
    \includegraphics[width=1\linewidth]{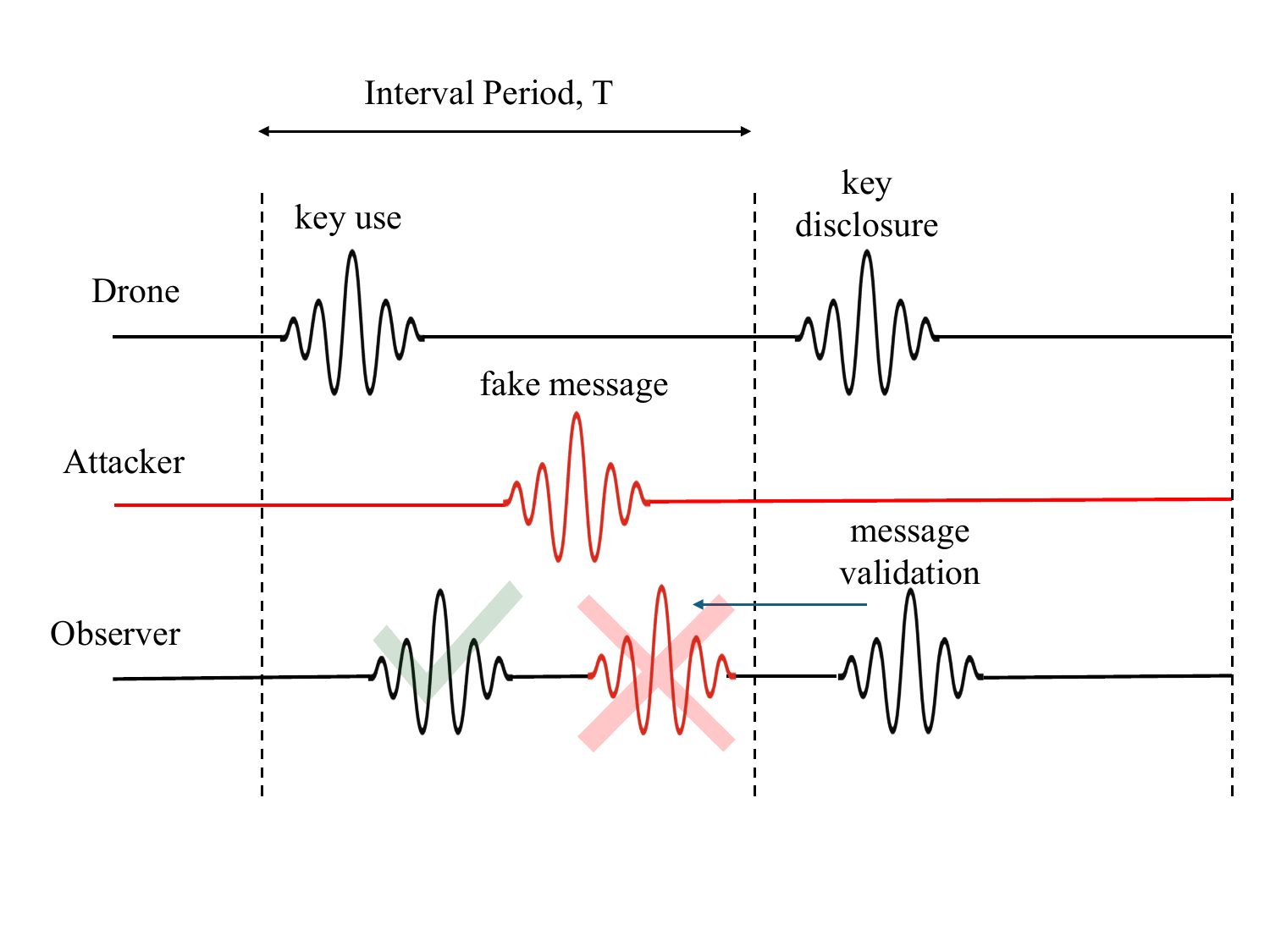}
    \caption{Fake Message Injection Within Key Interval. The attacker can inject a fake message during the current key interval, and forgery is detected by the observer in the subsequent interval.}
    \label{fig:fake_message}
    \Description[<short description>]{<long description>}
\end{figure}

TESLA requires loose time synchronization between transmitter and receiver to maintain security guarantees~\cite{perrig_tesla_2002}. When a receiver wishes to verify the authenticity of a received message, the primary goal is to ensure that the key used to calculate the HMAC of the payload was still a secret at the time of HMAC generation. In other words, the receiver must not trust a message that uses a key that has already been disclosed. Thus, time synchronization between the sender and receiver must be maintained to ensure all parties are in lock-step with when the key has been used and when a new key is disclosed. Our system assumes GPS time to synchronize the time between the transmitter and the receivers. Spoofing or jamming GPS as a mechanism to attack the TESLA protocol is beyond the scope of this work. However, we acknowledge that even with all parties using GPS synchronized clocks, there will still be some clock skew. One relevant work investigated the impacts of GPS clock synchronization on VANETs implementing TESLA authentication. It was determined that even with a maximum clock skew of 1.5 ms, there is potential for an attack if additional protections are not implemented ~\cite{jahanian_analysis_2015}. Beyond the clock skew introduced by GPS, special care must be taken to account for wireless broadcast protocols that may delay the transmitter's ability to broadcast messages within the intended interval. As we discussed in Section \ref{Implementation}, we propose using Wi-Fi beacons within a permitted transmission interval related to the interval period. This ensures that the media access control mechanisms imposed by the 802.11 protocol will not negatively affect the security guarantees of our proposed system. As each wireless broadcast mechanism employs a different media access control mechanism (e.g., Bluetooth, cellular), the parameters of TBRD must be modified to account for varying applications.

\textbf{Replay Attacks}- True replay attacks (i.e., re-broadcasting an old, valid message) are prevented by two mechanisms. First, the interval counter \(i\), included in every authentication payload prevents the re-use of a message within the same mission . Second, the USS mission time check ensures that an attacker cannot replay a captured broadcast from a previous, completed mission .

\textbf{UAS Capture or Compromise}- The goal of capturing a UAS by an attacker is to recover any key material from the UAS somehow. In the case of a system using asymmetric cryptography, the goal would be to recover the private signing key used to sign Remote ID broadcasts. This would significantly impact the system's security, as the asymmetric signing primitives are now compromised. Depending on the system's implementation, this could affect the integrity of past and future missions. However, since TBRD utilizes the TESLA protocol, the key material that an adversary could recover only affects the current active mission. Any past or future missions are intrinsically secure from this kind of attack. Since a UAS capture would be immediately evident to the UAS operator, a mechanism to inform the USS of a failed or invalid mission can be used to revoke the validity of the key commitment sent to the USS.

\textbf{Mobile Device Compromise} - A compromise of the mobile device used to generate the interval keys and report a key commitment to the USS is perhaps the most impactful attack on the overall security of TBRD. Should an attacker gain access to the key material used to generate the one-way key chain, the security guarantees of the TESLA protocol would no longer be upheld. This is why we have chosen to leverage a TEE to ensure the highest level of protection against attacks. Storing the most critical key material within the TEE ensures a higher level of resistance against attacks from the mobile device operating system or other malicious applications installed on the device. A future enhancement of the TBRD protocol could leverage additional forms of mobile identity management services, such as those provided by physical or embedded Subscriber Identity Modules (SIMs), to further enhance security.

\subsection{Computation and Communication Overhead}

\textbf{Computation Analysis:} To validate claims that the TBRD protocol provides reduced power consumption when compared to authentication through digital signatures, we developed an experimental method to examine the power necessary for signature generation as a measure of computational efficiency. As many UAS must operate under strict battery power limitations, power consumption is a key concern. \
For our analysis, we must ensure that we compare algorithms that provide a similar level of security. While the security strength of cryptographic algorithms depends on many factors, we can rely on well-established algorithms with clear standards to ensure a systematic approach. One such standard is NIST Special Publication 800-57 Part 1~\cite{barker_recommendation_2020}. This standard compares various cryptographic algorithms and assigns a security strength (measured in bits) as a function of the algorithm type and key size.  Our implementation of the TBRD protocol utilizes a SHA-256 HMAC to achieve authentication, providing 256 bits of security strength. To provide an equivalent level of security (256 bits), an elliptic-curve cryptography (ECC) key size of at least 512 bits is required~\cite{barker_recommendation_2020}.

We chose to conduct our tests of cryptographic power consumption using the ESP32-S3-DevKitM-1~\cite{espressif_esp32-s3-devkitm-1_2025}. This platform was chosen due to its ease of use, support for rapid prototyping and testing, as well as the prevalence of ESP32-based Remote ID transponders already in everyday use.  As mentioned previously, ArduPilot has spearheaded the open-source implementation of a Remote ID-compliant transmitter. The list of supported hardware includes seven boards, including the ESP32-S3 dev board, and several other ESP32-based solutions~\cite{ardupilot_ardupilot_2025, holybro_remote_2025,blue_mark_dronebeacon_2025}. The boards offer a compact, lightweight form factor for UAS operators to retrofit existing UAS platforms with Remote ID compliant transmitters.

\begin{figure}
    \centering
    \includegraphics[width=1\linewidth]{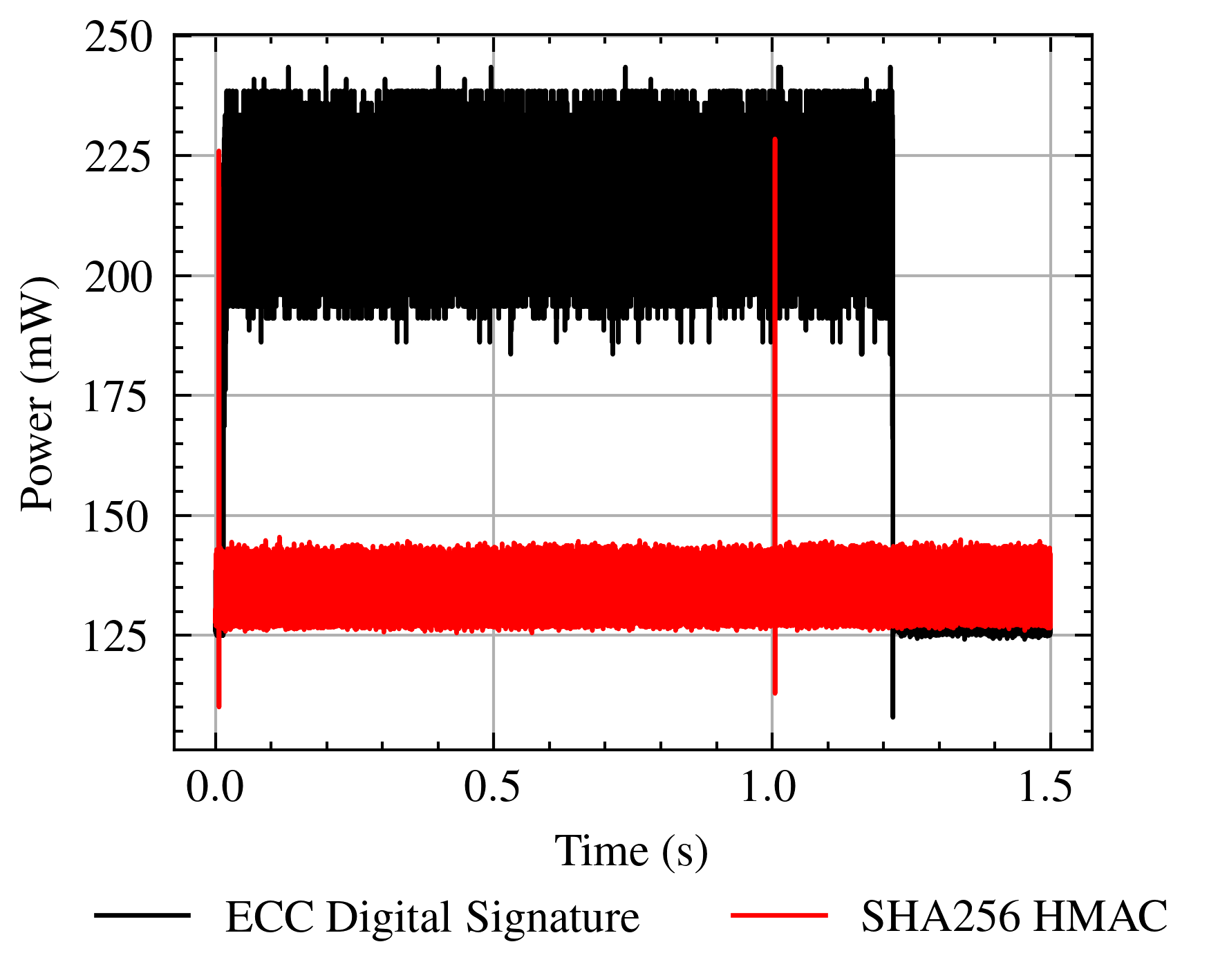}
    \caption{Power consumption of an ESP32-S3 microcontroller computing the ECC digital signature of a 100-byte payload compared to the same device computing a SHA256 HMAC.}
    \label{fig:Power}
    \Description[<short description>]{<long description>}
\end{figure}

To isolate the effects of the message authentication process, we programmed our ESP32-based microcontrollers to calculate a SHA256 HMAC of a 100-byte payload once every second using the WolfSSL embedded cryptography library~\cite{wolfssl_wolfssl_2017}. The 100-byte payload mirrors the dynamic flight data, used to generate the HMAC used by the TBRD protocol, and the signature rate complies with the ASTM standard broadcast rate for Remote ID. Using a Nordic Semiconductor Power Profiler II device, we were then able to measure both the computation time and the amount of power used during the HMAC calculations \cite{nordic_semiconductors_power_2025}.  Similarly, we can analyze the calculation of an ECC digital signature for a 100-byte payload with a 512-bit key size.  However, the ESP32-S3 was unable to compute the digital signature at a rate of once per second.  While other microcontrollers may be able to compute the digital signature at a faster rate, the configuration we tested would not be able to comply with the minimum transmission rate specified by the Open Drone ID standard~\cite{segger_emsecure-ecdsa_2025}. Figure \ref{fig:Power} depicts the power consumption used by the microcontroller during computational intervals. On average, the SHA256 HMAC took 10 ms while the ECC digital signature took 1.2 seconds. Additionally, it is worth noting that the SHA256 HMAC generates a 32-byte hash output, whereas the ECC algorithm produces a 139-byte signature.  The average power required for a SHA256 HMAC is 129 mW, while the average power needed for an ECC digital signature is 216 mW.

\textbf{Communication Overhead:} There are several constraints we must assess to evaluate the performance of the TBRD protocol.  Since we have chosen to implement TBRD within the framework of the ASTM Open Drone ID standard, we must ensure that the additional data for authentication conforms to the relevant message size constraints. The standard limits the total length of all authentication/signature data to 255 bytes. The TBRD authentication data is 68 bytes in length when using a SHA-256 HMAC for authentication. It is essential to note that the authentication data for TBRD may be compromised with trade-offs to system security. The SHA256 HMAC provided 256 bits of security, while the TESLA protocol used by Galileo operates with a truncated HMAC of only 40 bits~\cite{european_spce_agency_galileo_2021}. If we compare the use of an equally secure digital signature scheme like ECC, the signature size would double the necessary length of the authentication data. When operating in a wireless environment with a high potential for packet loss, it is crucial to minimize the overhead required for message authentication. TBRD excels in this regard as the same message size can be authenticated with a smaller amount of data compared to a digital signature approach.

\section{Discussion} \label{Results}
%\textbf{TBRD Parameter Configuration:} The TBRD protocol allows for many adjustments to system parameters to enable trade-offs in security and performance. The parameters include the decision in the delay disclosure period, the interval period, and the adjustable HMAC size.  The TESLA protocol introduces a controlled delay \(d\) between the time a message is sent and when the authentication key is disclosed. This delay must be carefully balanced against real-time requirements. In latency-sensitive applications (e.g., drone command and control), a long delay may degrade responsiveness, while a shorter delay could reduce the time available for key distribution and verification. The interval duration \(T_{int}\) must also be tuned to match message transmission frequency and expected network conditions. For example, high message frequency may require shorter \(T_{int}\) to minimize overlap and reduce buffer size requirements. Conversely, longer intervals may reduce overhead but increase exposure to packet loss or adversarial manipulation. Careful parameter selection is crucial for optimizing both security and operational performance. Lastly, the HMAC output may be truncated to address constraints resulting from even more lossy wireless communication channels.

\textbf{TBRD Parameter Configuration:} The TBRD protocol supports several adjustable parameters to balance security and performance, including the delay disclosure period, interval period, and HMAC size. TESLA introduces a controlled delay \(d\) between message transmission and key disclosure, which must be balanced against real-time needs. In latency-sensitive applications (e.g., drone command and control), longer delays degrade responsiveness, while shorter ones limit time for key distribution and verification. The interval duration \(T_{int}\) should align with message frequency and network conditions: high-frequency messaging may require shorter \(T_{int}\) to reduce overlap and buffer demand, while longer intervals reduce overhead but increase exposure to packet loss or manipulation. Optimal parameter tuning is crucial for striking a balance between security and operational efficiency. The HMAC output may also be truncated to accommodate lossy wireless channels.

\textbf{Scalability:} Any message authentication protocol must address the impacts of scalability. The ability to accomplish broadcast authentication is influenced by the robustness of the underlying communication channel as the number of transmitters and receivers grows. As the number of UAS platforms transmitting in close proximity grows, the potential for packet loss grows, particularly in RF-congested or mobile environments. Since TESLA requires both the message and the corresponding disclosed key to be received and matched within a bounded time window, packet loss can result in failed authentication and the loss of legitimate messages. This issue becomes more pronounced in lossy wireless environments (e.g., Wi-Fi or Bluetooth advertisements), where retransmission is not guaranteed. The TESLA broadcast authentication protocol excels in accommodating large numbers of potential unknown receivers. 

\textbf{Mobile compromise:} In TBRD, the key commitment of the TESLA-based broadcast authentication depends on the mobile device to serve as a trusted execution platform for key disclosure and signature operations. However, the integrity of this process hinges on the security guarantees of the TEE itself. Current TEEs (e.g., ARM TrustZone or Apple Secure Enclave) offer limited visibility and control to developers, and vendor-specific APIs or OS policies often constrain their use. Moreover, if the mobile device is compromised (e.g., rooted or jailbroken), the adversary may gain access to pre-disclosure keys. This highlights a significant limitation: the security model assumes the mobile TEE is trustworthy, but practical deployments must account for the possibility of device-level compromise.

\textbf{Key reuse detection and revocation:} TESLA relies on the time-ordered disclosure of one-time keys. However, key reuse — whether accidental or malicious — undermines the protocol’s security guarantees by enabling replay or forgery. TBRD presents detection mechanisms to identify repeated use of the same key for different messages or across different intervals.

\textbf{Standards integration:} Integrating TESLA into the ASTM Remote ID framework presents both opportunities and challenges. TESLA offers a lightweight, broadcast-friendly authentication primitive that complements the ASTM requirement for message authenticity without necessitating public key infrastructure on resource-constrained drones. However, compliance with current ASTM standards (e.g., F3411) requires strict adherence to message construction and transmission. Furthermore, compatibility with various transmission mechanisms (Bluetooth, Wi-Fi) adds additional constraints.

\section{Related Work} \label{Related}
Several works have focused on the privacy and security issues associated with compliance with the Remote ID rule. In ~\cite{tedeschi_privacy-aware_2024}, Tedeschi et al. provide perhaps the most comprehensive overview of privacy and security issues facing Remote ID compliance by UAS operators and manufacturers. Their work identifies several potential threats and solutions, considering the security, privacy, and energy trade-offs associated with these solutions. While a wide variety of technical approaches were considered, their work did not address or mention the use of the TESLA algorithm.

Of the numerous works that propose a technical solution or protocol to enhance Remote ID security, the most closely related is CAPSID~\cite{li_capsid_2024}. In their work, Li et al. propose the CAPSID protocol, which builds upon the proposed use of session IDs. At its core, CAPSID utilizes certificates to authenticate broadcast messages. Our solution leverages the TESLA protocol to achieve authentication in a more computationally efficient manner. Additionally, they do not provide a practical solution to UAS capture. TBRD ensures that capture of a UAS does not give an adversary access to long-term key material.  Furthermore, other closely related works attempt to provide solutions that enhance operator privacy. Privacy of operator location was not a focus of our work~\cite{li_capsid_2024,tedeschi_arid_2021,wisse_2_2023}.

The application of TESLA to solve current and future issues facing broadcast authentication is not unique. TESLA is in use by the Galileo global navigation satellite sytem~\cite{european_gnss_supervisory_authority_galileo_2022}. TESLA was proposed as a solution for providing authentication to ADS-B in \cite{berthier_sat_2017}, which is widely used in the aviation industry. For the maritime sector, it has been proposed to enhance the authentication of Automated Identification Systems (AIS)~\cite{sciancalepore_auth-ais_2022} as well as its replacement, VHF Data Exchange (VDES) ~\cite{wimpenny_pragmatic_2025}. Our system has many similarities with AuthAIS. One key difference is that AuthAIS utilizes a Trusted Third Party Maritime Authority to assign and distribute root keys to vessels that wish to use TESLA-authenticated AIS. They also propose a probabilistic security framework that reduces message overhead. Using TESLA to authenticate messages in these industries has not yet gained widespread adoption among the mainstream. The rapid expansion and constraints of UAS operations demand new solutions; our approach unites TESLA and trusted computing to surpass earlier limitations.

\section{Conclusion} \label{Conclusion}
This paper presents TBRD, a system to enhance the security of UAS operations by providing a means of authenticating broadcast Remote ID. While solutions that implement digital signatures to achieve authentication might seem viable, communication bandwidth and computational time constraints are significant detractors. We produce the first fully implemented solution that combines the security guarantees offered by secure mobile computing with the symmetric cryptography-based TESLA broadcast authentication algorithm.  We designed and implemented TBRD to protect against a wide array of threats, including spoofing, record and reply, and UAS capture. Furthermore, the system was evaluated using state-of-the-art simulation techniques to prove system reliability. The need to secure the airspace against threats from unauthorized UAS operations is greater now than ever. Systems like TBRD must be adopted to protect public safety and national security.

\bibliographystyle{ACM-Reference-Format}
\bibliography{references}

\end{document}